%% file: main.tex
\documentclass[journal]{vgtc}                

\ifpdf
  \pdfoutput=1\relax                   
  \usepackage{graphicx}                
  \DeclareGraphicsExtensions{.pdf,.png,.jpg,.jpeg} 
\else
  \usepackage{graphicx}                
  \DeclareGraphicsExtensions{.eps}     
\fi%

\graphicspath{{figures/}{pictures/}{images/}{./}} 

\usepackage{microtype}                 
\PassOptionsToPackage{warn}{textcomp}  
\usepackage{textcomp}                  
\usepackage{mathptmx}                  
\usepackage{times} 

\usepackage{cite}                      
\usepackage{algorithmic}
\usepackage{tabu}                      
\usepackage{booktabs}                  
\usepackage{amsmath}

\usepackage{mathtools}
\DeclarePairedDelimiter{\ceil}{\lceil}{\rceil}

\usepackage{balance}  
\usepackage{graphics} 
\usepackage{times}    
\usepackage{url}      
\usepackage[ruled,vlined]{algorithm2e}
\usepackage{enumitem}
\usepackage{xspace}
\usepackage[table,usenames,dvipsnames]{xcolor}
\usepackage{booktabs}
\usepackage{multirow}
\usepackage[10pt]{moresize} 
\usepackage{changepage}
\usepackage[mathscr]{euscript}
\usepackage{amssymb}

\usepackage{algorithmic,float}
\usepackage{tikz}

\onlineid{1073}

\vgtccategory{Research}
\vgtcpapertype{application/system}

\definecolor{rvtext}{RGB}{84,126,201}

\newcommand{\rv}[1]{{\leavevmode\color{black}{#1}}}

\newcommand{\name}{CloudDet\xspace}
\newcommand{\experts}{experts\xspace}
\newcommand{\ea}{\textbf{E1}\xspace}
\newcommand{\eb}{\textbf{E2}\xspace}

\title{\name: Interactive Visual Analysis of Anomalous Performances \\in Cloud Computing Systems}

\author{Ke Xu, Yun Wang, Leni Yang, Yifang Wang, Bo Qiao, Si Qin, Yong Xu, Haidong Zhang, Huamin Qu}
\authorfooter{
\item
 Ke Xu, Leni Yang, and Yifang Wang are with the Hong Kong University of Science and Technology. E-mail: \{kxuak, lyangbb, ywangjh\} @connect.ust.hk. This work is done when Ke Xu is an intern at MSRA.
\item
 Yun Wang, Bo Qiao, Si Qin, Yong Xu, and Haidong Zhang are with Microsoft Research. E-mail: \{wangyun, boqiao, Si.Qin, yox, haizhang\} @microsoft.com. Yun Wang is the corresponding author.
\item
 Huamin Qu is with the Hong Kong University of Science and Technology. Email: huamin@cse.ust.hk.
}

\input sections/00-abstract


\teaser{
  \centering
  \includegraphics[width=\linewidth]{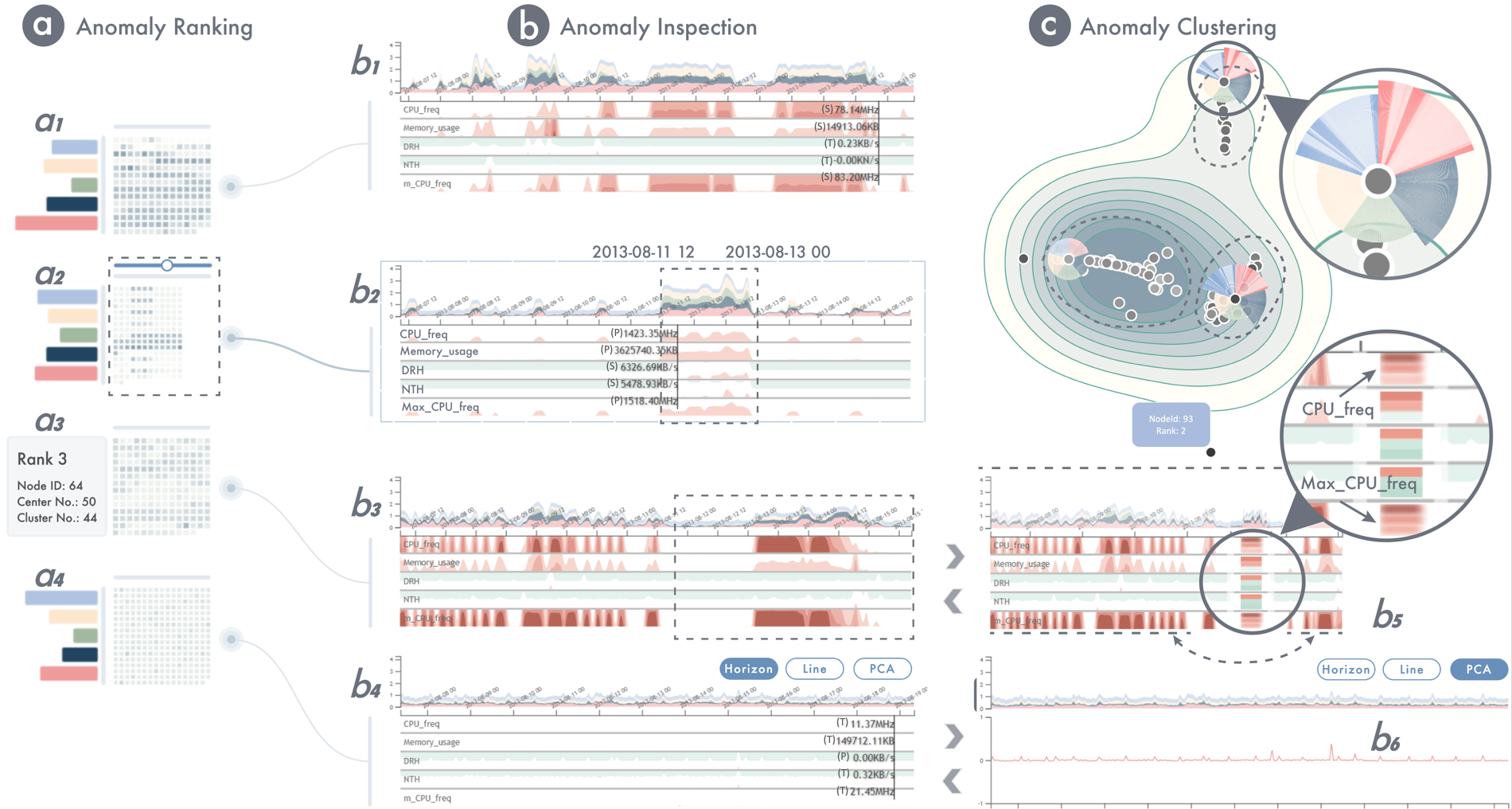}
  \caption{\name facilitates the exploration of anomalous cloud computing performances through three levels of analysis: (a) anomaly ranking, (b) anomaly inspection, and (c) anomaly clustering. The figure showcases some exploration results with Bitbrains Datacenter traces data. Node (b1) contains both short and long term spikes, with no pattern in their occurrence times. Node (b2) shows a 12-hour periodic pattern for the performance metrics by observing the calendar chart in (a2), but encounters a spike in the process. Node (b3) shows many short-term and near-periodic spikes at the beginning and an abnormal long-term spike near the end. After collapsing the long-term one into a visual aggregation glyph in (b5), (b3) is updated and the latter temporal data ``pop out'', which shows a similar pattern as the beginning. Node (b4) shows a general periodic trend which is not apparent in (a4) by using the PCA analysis in (b6). Most of the nodes are clustered into three groups in (c), with each group displaying a rare but similar performance.}
  \label{fig:case1}
}

\vgtcinsertpkg

\begin{document}

\input sections/01-intro

\input sections/02-related

\input sections/03-overview

\input sections/04-algorithm

\input sections/05-visual

\input sections/06-evaluation

\input sections/07-conclusion

\input sections/09-acknowledgement


\bibliographystyle{abbrv-doi}

\bibliography{main}
\end{document}

%% file: sections/00-abstract.tex
\abstract{
Detecting and analyzing potential anomalous performances in cloud computing systems is essential for avoiding losses to customers and ensuring the efficient operation of the systems. To this end, a variety of automated techniques have been developed to identify anomalies in cloud computing performance. These techniques are usually adopted to track the performance metrics of the system (e.g., CPU, memory, and disk I/O), represented by a multivariate time series. However, given the complex characteristics of cloud computing data, the effectiveness of these automated methods is affected. \rv{Thus,} substantial human judgment on the automated analysis results is required for anomaly interpretation. In this paper, we present a unified visual analytics system named \name to interactively detect, inspect, and diagnose anomalies in cloud computing systems. A novel unsupervised anomaly detection algorithm is developed to identify anomalies based on the specific temporal patterns of the given metrics data (e.g., the periodic pattern), the results of which are visualized in our system to indicate the occurrences of anomalies. Rich visualization and interaction designs are used to help understand the anomalies in the spatial and temporal context. We demonstrate the effectiveness of \name through a quantitative evaluation, two case studies with real-world data, and interviews with domain experts.}  

%% file: sections/01-intro.tex
\firstsection{Introduction}
\label{sec:intro}
\maketitle
\vspace{-0.05cm}
Cloud computing is becoming increasingly pervasive, with the extensive demand for big data analytics and discovery shifting many individuals and organizations towards cloud services. This move is motivated by benefits such as shared storage and computation service among a massive number of users. In order to maximally leverage the cloud, high availability and reliability are of utmost importance to the overall user experience. 
Therefore, it is important to monitor the compute nodes' usage and behavior, and then gain insights into the potential anomalous operations running in the cloud which might result in reduced efficiency or even downtime of the data center. 

The most efficient way to detect and analyze anomalies in cloud systems is to monitor the general performance metrics of compute nodes (servers or virtual machines (VMs)), including CPU, memory, disk I/O, etc., in the form of a set of multivariate time series \cite{muelder2016visual}. While monitoring applications hosted by these nodes also makes sense, the efficiency of this method is hindered by the scale, privacy and noise issues. Thus\rv{,} tracking the performance metrics of compute nodes is more reasonable for exploring unusual behaviors. However, it is difficult to make automated anomaly detection with performance data in cloud systems. First, traditional methods for anomaly detection have mainly been approached through statistical and machine learning techniques \cite{chandola2009anomaly, hodge2004survey, sari2015review}, 
which face the inherent limitation that the ground truth required for training and performance evaluation is usually difficult to obtain. Second, the performance data generated by cloud systems are unstructured, and of high velocity. A very large-scale cloud data center often contains thousands of compute nodes that can record different metrics in a minute granularity. Third, the automated methods usually generate too many false positive results that aggravate the diagnosis burden in real-world scenarios. 
These characteristics stress the need for a more scalable and flexible way to identify and interpret anomalies in cloud computing.

Data visualization facilitates the analysis and evaluation of anomaly detection results via rich interaction techniques and intuitive representations of contextual information that provides
additional evidences to support or refute the analysis conclusions \cite{xie2019visual, akoglu2015graph, mckenna2016bubblenet}. However, most existing solutions are not specifically designed for anomaly detection in cloud computing, or are limited in capability for exploring large-scale multivariate time-series data. The main challenges challengess include: (1) Scalability: given the scale of compute nodes in cloud systems,
there is a need to clearly show the anomalous patterns and to optimize the trade-off between system scalability and level-of-detail (LoD) of the performance data. (2) Interpretability: it is difficult to intuitively represent anomalous patterns and their corresponding raw data context with different semantics. (3) Multi-dimensionality: to comprehensively understand the data and anomalies, multi-faceted patterns should be conveyed, such as the temporal patterns of a compute node in terms of 
multiple metrics (e.g., the correlation between CPU and memory usage), or the distribution patterns among different nodes.

To address the challenges, we introduce \name, an interactive visualization system to \rv{help cloud service providers efficiently detect and diagnose anomalies in large-scale cloud computing systems with the system-level performance data.} Major research contributions include:

\begin{itemize}[leftmargin=*,topsep=4pt]
    \vspace{-2mm}
     \item \textbf{System.} We propose an integrated visual analytics system for the interactive exploration, detection, and diagnosis of anomalies with large-scale performance metric data in cloud computing systems. We formulate the design requirements through cooperation with cloud operations engineers. An unsupervised algorithm is proposed to \rv{facilitate the evaluation of anomalies based on the captured change patterns in a time series with ensemble analysis.} The visualization and interaction designs support the visual analysis of anomalies at three levels: anomaly overview, ranking, and diagnosis.

    \vspace{-2mm}
    \item \textbf{Visualization and Interactions.} We propose a set of visualization and interaction designs to facilitate users' \rv{ranking,} inspection and perception of most abnormal performances in cloud computing data. Specifically, we extend horizon graphs to \rv{make it visually scalable} for displaying the overall pattern and making comparisons of different metrics. A glyph based on visual aggregation technique is introduced in the horizon graph to support a nonlinear time scaling. Another glyph is designed for data abstraction and anomaly highlighting in the spatial distribution context. Rich interaction designs are used in different views to promote data exploration.

    \vspace{-2mm}
    \item \textbf{Evaluation.} The effectiveness of \name is demonstrated in multiple forms of evaluation. We first conduct a quantitative performance evaluation of the anomaly detection algorithm. \rv{The results show that our algorithm outperforms baseline algorithms in terms of accuracy with a comparable scalability.} We then describe how \name works through two case studies with real-world datasets, and collect feedback from experts in the cloud computing domain. \rv{All the evaluations validate the effectiveness of \name in analyzing anomalous performances with cloud computing data.}

\end{itemize}

%% file: sections/02-related.tex
\section{Related Work}
\label{sec:related}

In this section, we summarize the techniques that are most related to our work, including anomaly detection algorithms and visualizations for anomaly detection and temporal data.

\subsection{Anomaly Detection}
Anomaly detection has been extensively studied in recent years. 
Existing methodologies include classification-based algorithms 
(either supervised \cite{hawkins2002outlier, mahoney2003learning, wong2003bayesian} or semi-supervised \cite{chen2001one, liu2008isolation})
, statistics-based algorithms \cite{barnett1974outliers, yamanishi2004line}, distance-based algorithms \cite{hautamaki2004outlier, breunig2000lof, bay2003mining}, and spectral-based algorithms \cite{shyu2003novel}.
Different approaches have been taken to address the problems with time-varying, multivariate data \cite{gupta2014outlier}, which is the type of performance metrics data in cloud computing systems. For example, one major category employs regression model-based algorithms to determine the anomaly score of time-series data \cite{abraham1989outlier,fox1972outliers}, and these have been extended to many subsequent models like the autoregressive moving average (ARMA) and autoregressive integrated moving average (ARIMA). 
\rv{Although these regression-based algorithms are useful to deal with most applications with various data types, they have a limited capability in cloud computing systems due to the huge volume and variation of data.}

A variety of anomaly detection techniques have been specifically designed for performance anomaly detection in cloud computing systems \cite{ibidunmoye2015performance, sari2015review, agrawal2017adaptive, wang2016fd4c}. 
Some approaches deal with specific types of problems in cloud systems. For example, Xu et al. focused on the anomalies during the execution of sporadic operations \cite{xu2014pod}, and Yu et al. presented a diagnosis framework for Hadoop environment \cite{yu2013scalable}. More general algorithms without domain-specific assumptions have also been introduced recently, such as techniques developed based on the statistical techniques \cite{wang2011statistical, wang2010online}, entropy-based models \cite{navaz2013entropy} and isolation-based algorithms \cite{calheiros2017effectiveness}. However, most of them are computationally
. By contrast, scalable algorithms have also been proposed to facilitate the anomaly detection in cloud computing, e.g., 
implementing a probabilistic approach to detect abnormal software systems \cite{shen2009reference}, adopting Holt-Winters forecasting to identify a violation in application metrics \cite{jehangiri2014diagnosing}, and implementing a clustering method to find the anomalous application threads \cite{zhang2016taskinsight}. A common issue of these \rv{scalable techniques is that they require low-level access to application level information, while our approach only targets general performance metrics that can be obtained via sampling the state of the system.} Another crucial limitation of these approaches is that \rv{many of them are specified for different anomaly patterns in data \cite{xu2014pod, calheiros2017effectiveness}}, lacking a systematic and unified way to detect anomalies based on data features and to interpret the results derived from these techniques.

In this paper, we design an unsupervised algorithm that uses different strategies to evaluate anomalous performances based on their change patterns. The algorithm is also integrated in our system to help interpret normal and abnormal cases situated in cloud performance data.

\subsection{Visualization for Anomaly Detection}
Although the aforementioned algorithms generate numeric results of anomalies, they are limited in providing interpretation of the anomalies. The problem is further aggravated when there is a lack of a clear boundary and ground truth for normal/abnormal cases for evaluation. Therefore, the domain knowledge of human experts need to be involved in judging the cases. Visualization techniques have been applied to support interpretation and facilitate better decision making. In the traditional approach, statistical diagrams such as line charts and histograms are commonly used to represent anomalous trends in the raw data
\cite{kind2009histogram, laskov2005visualization, lin2005visualizing}. A variety of dimensionality reduction (DR) techniques are also applied for understanding how the data distribute in a multidimensional feature space, such as multidimensional scaling (MDS) \cite{kruskal1964multidimensional}, principal component analysis (PCA) \cite{shyu2003novel} and t-distributed stochastic neighbor embedding (t-SNE) \cite{maaten2008visualizing}. However, a crucial limitation is that \rv{these DR methods} have a narrow focus that can not provide a systematic approach for other complicated applications. 

Recently, more-unified visual analytics systems for anomaly detection have been proposed. For example, systems have been developed for temporal, multivariate data \cite{tao2018visual, xu2018ecglens}, for applications such as monitoring streaming traffic data \cite{cao2018voila}, detecting the spreading of rumors \cite{zhao2014fluxflow}.
More specific visualizations for performance analysis in cloud computing systems have also been introduced \cite{peiris2014pad, farshchi2015experience}. In a very recent work, Cong et al. \cite{xie2019visual} proposed a visual analytic approach to detect anomalous executions with their call stack tree (CSTree) representation in high performance computing applications. However, \rv{it detects anomalies in the cloud system with event sequence data, like the log events, which is application-level analysis rather than the general system-level performance analysis}. 

In our work, we analyze anomalous compute nodes by tracking the general, time-varying performance metrics (profile data) in cloud computing systems, such as CPU load, memory usage, and disk read/write load, etc. Novel visualization and interaction techniques are also introduced to help users interactively identify the anomalies by inspecting different performance metrics and the correlations among them.

\subsection{Visualization for Temporal Data}
The performance data tracked in cloud computing systems are represented by multivariate time series.
Given the ubiquity of temporal data, researchers have broadly studied their visualization as applied to various applications \cite{shneiderman2003eyes}. Several surveys reported the state-of-the-art multivariate time series visualization techniques. M{\"u}ller et al. \cite{muller2003visualization} discussed the general aspects of time-dependent multivariate datasets 
and categorize the visualization techniques into static representations and dynamic representations. Aigner et al. \cite{aigner2008visual} summarized the visual methods for analyzing temporal data from three aspects: temporal data abstraction, principal component analysis (PCA), and clustering. 

Among the myriad approaches to temporal data visualization, multivariate time series are most relevant to our work. These approaches can be classed into four types \cite{isaacs2014state, ezzati2017multi}: (1) Line charts are the most common method to visualize the progression of values over time.
Recently, Muelder et al. \cite{muelder2016visual} designed behavioral lines to analyze the behavior of cloud computing systems by bundling different performance metrics over time \cite{muelder2016visual}. However, placing multiple lines in the same space will reduce the fidelity of individual time series. (2) Stacked graphs are another approach which provides a summation view stratified by individual series \cite{byron2008stacked}. Projects like NameVoyager \cite{wattenberg2006designing} and sense.us \cite{heer2007voyagers} applied stacked graph to explore demographic data. 
The problems with stacked graph include difficulty in comparing individual series and misinterpretation of the space between curves. (3) Horizon charts are a space-efficient technique for time series visualization.
Such charts were first proposed by Saito et al. \cite{saito2005two}, and then optimized in terms of graphical scalability and perception by Few et al.\cite{fewtime} and Heer et al. \cite{heer2009sizing}. (4) More recently, many glyph-based designs have been developed to visualize the time series data in various applications \cite{fuchs2013evaluation}. For example,
TargetVue \cite{cao2016targetvue} introduced three different glyph designs to detect anomalous user behaviors in their communication activities. Glyphs are an appropriate choice due to their effective use of screen space and expressiveness for temporal performance summary.

In our work, we enhance horizon charts by combining a glyph design that can visually aggregate the performance data to support the visual scalability in the time domain, thus\rv{,} facilitating the comparison of focused data and the identification of anomalous nodes in large-scale cloud computing systems. Another glyph is designed to help the anomaly comparison based on the similarity of compute nodes. Various interaction techniques are introduced in our system to facilitate the exploration of multivariate time series from cloud computing systems.

%% file: sections/03-overview.tex
\section{System Overview}
\label{sec:overview}
\name was developed as part of a one-year anomaly detection project. This project is aimed at addressing the aforementioned challenges (scalability, interpretability, and multi-dimensionality) and satisfying real-world requirements formulated by \rv{three domain experts in anomaly detection and cloud system analysis. Weekly meetings were held for two months, at which detailed system design requirements were clarified by the discussion of anomaly detection problems in both cloud computing and visualization}. One pilot system was developed to provide a rough evaluation of the anomaly detection algorithms. Below we describe the most critical design requirements (\textbf{R1--R4}) developed during these discussions.

\begin{figure} [t]
 \centering 
 \includegraphics[width=\linewidth]{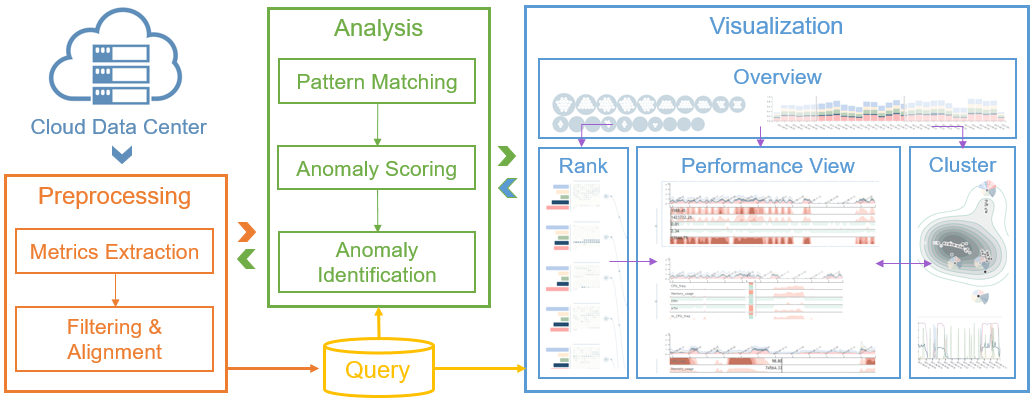}
 \vspace{-0.4cm}
 \caption{\name system overview and data processing pipeline.}
 \vspace{-0.4cm}
 \label{fig:system}
\end{figure}

\begin{enumerate}[leftmargin=*,label={\textbf{R{\arabic*}}}]
 \setlength{\itemsep}{2pt}
 \setlength{\parskip}{2pt}
 \setlength{\parsep}{2pt}

\item \textbf{Support large-scale cloud computing data.}\rv{
The system should allow effective analysis and exploration of large-scale cloud performance data, enabling users to examine the anomalous cloud computing behaviors collected from thousands of compute nodes.}

\item \textbf{Facilitate anomaly detection processes.}
Automatic anomaly detection algorithm should be integrated into the system to support the efficient identification of anomalies in historical data.The visual encoding should efficiently rank and highlight the abnormal nodes, as well as show when and how the anomalies occurred. 

\item \textbf{Enhance the anomaly inspection with multifaceted pattern discovery.} 
The system should create easy-to-understand designs to connect the anomaly detection results with auxiliary information from the raw performance data to form a semantic background against anomaly instances. 

\item \textbf{Enable users to explore the data interactively.}
\rv{To facilitate expert users analyzing the detection results, it is necessary to incorporate flexible interactions that help them quickly explore a substantial number of performance tracking data.} 

\end{enumerate}

\begin{figure*} [t]
 \centering 
 \includegraphics[width=\linewidth]{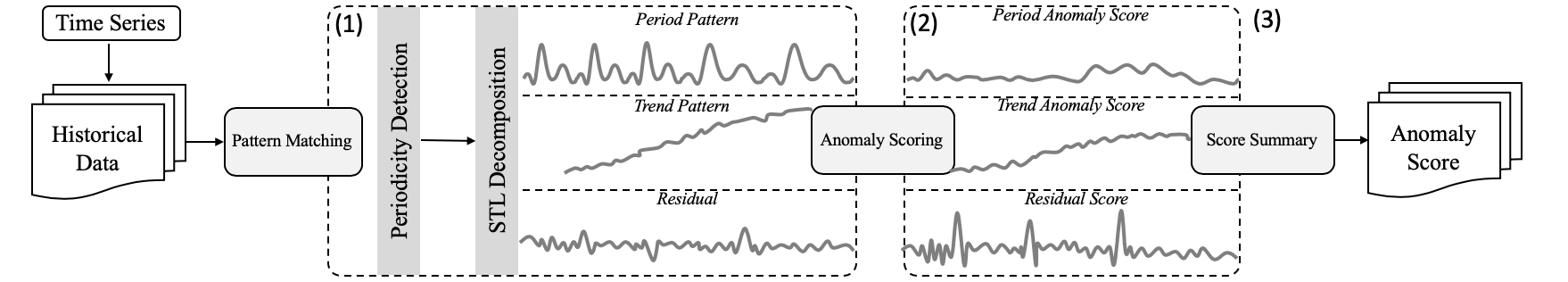}
 \vspace{-0.6cm}
 \caption{Algorithm pipeline: (1) pattern matching for time series, (2) anomaly scoring, and (3) anomaly score summary.}
 \vspace{-0.4cm}
 \label{fig:algorithm}
\end{figure*}

Based on these requirements, we have designed \name, an interactive data exploration system that enables \experts to visually identify and analyze the anomalies in cloud computing performance. 
Fig. \ref{fig:system} illustrates the system architecture and the interactive analysis processing pipeline. The system consists of four major steps: (1) the data collection module, (2) the data preprocessing module, (3) the analysis module and (4) the visualization and interaction module. In particular, the data collection module collects the large-scale cloud computing performance data (\textbf{R1}). The data processing runs on the database, \rv{where different compute nodes' performance metrics are aligned and transformed into time-series data for anomaly scoring (\textbf{R2}). The analysis module runs an unsupervised anomaly detection algorithm to detect suspicious nodes and their abnormal periods for different metrics according to their temporal patterns, and ranks these nodes based on their anomaly scores (\textbf{R2}).}
The visualization and interaction module displays anomalous nodes, together with the corresponding contexts of raw metric data within several views. \rv{It provides a comprehensive visual summarization and interpretation of different patterns in cloud computing performance (\textbf{R3}). Rich interaction designs are created to support efficient anomaly exploration and diagnosis (\textbf{R4}). }All these modules work together to form a scalable mechanism that enables an efficient procedure to reduce the information seeking space.

%% file: sections/04-algorithm.tex
\section{Time Series Anomaly Detection}
\label{sec:analysis}

In this section, we introduce a novel unified algorithm that detects anomalies in time series based on the three important anomaly patterns from a business perspective (Fig.~\ref{fig:algorithm}). \rv{The three components are numerically represented by integrating the periodicity detection results with STL decomposition. Then we use an ensemble-based method to to calculate the anomaly score of time-series data with respect to the three patterns in the data.} The key procedures are described below.

\subsection{Pattern Matching}

\rv{The input time series, which are the metric tracking data of each compute node, are independently transformed into a set of historical data: for every data point $d_n$ in the time series ($n \geq L$), the previous $L$ points construct the temporal data $\textbf{D} = \{d_{n-L},d_{n-(L-1)},...,d_{n-1},d_n\}$, where $L$ is \rv{a} user-defined parameter, as the number of recent historical data.}
Then the historical data at different points are fed into the pattern matching module, which contains the two following steps.

\textbf{Periodicity Detection.} \rv{As shown in Fig.~\ref{fig:algorithm}(1), the first step is to detect the structural periodic changes and estimate the periodicity for each piece of historical data, which is an important parameter for the following steps. Here we employ a non-parametric, two-tier approach that provides both a high resolution and low false positive rate for periodic identification \cite{vlachos2005periodicity}. Specifically, the periodogram is firstly used to extract the candidate period, which is formulated as
  \vspace{-0.1cm}
 \begin{equation}
  \vspace{-0.1cm}
     P(f_{k/N}) = ||X(f_{k/N})||^2, ~~k = 0, 1... \ceil{\frac{N - 1}{2}},
 \end{equation}
where $P$ is the periodogram, indicating the expected power at each frequency $f_{k/N}$ (or equivalently at period $N/k$) of the input time series. The larger it goes, the more possible $N/k$ is the candidate periodicity. $X(f_{k/N})$ is the Discrete Fourier Transform of the input time series at frequency $f_{k/N}$. However, the periodogram might provide false candidates or coarse estimation of the period due to the discrete increasing of frequency $k/N$. Therefore, a verification phase, autocorrelation (ACF), is required to make a more fine-grained estimation of the potential periodicities. ACF can examine the similarity of a time series to its previous values for different $\tau$ lags by calculating the following convolution:
  \vspace{-0.1cm}
 \begin{equation}
  \vspace{-0.1cm}
     ACF(\tau) = \frac{1}{N}\sum_{n=0}^{N-1}d(\tau)\cdot{d(n+\tau)},~~\tau = 0, 1... \ceil{\frac{N - 1}{2}}.
 \end{equation}
If the candidate period from the periodogram lies on a hill (i.e., the local maximum) of ACF, we consider it as a valid period. Thus, for the historical data at the $n$th timestamp, we obtain the closest estimation of the period, denoted as $T_n$, which will be used as the parameter in the next step, and for anomaly scoring at the last step. \footnote{\protect\url{https://lukeluker.github.io/supp/PeriodicityDetection.py}}}

\textbf{STL Decomposition.} The historical data is further split into three components by the seasonal-trend decomposition procedure based on loess (STL) \cite{cleveland1990stl}, \rv{which is aimed to extract three temporal components for the anomaly score calculation. STL is proved to be fast and can cover most anomaly patterns in temporal cloud computing data. In addition, the three derived components are complementary in change patterns, and consistent in their unit and interpretability.} We use the additive model of STL:
  \vspace{-0.1cm}
 \begin{equation}
  \vspace{-0.1cm}
     d_n = S_n + Tr_n + R_n, ~~n = 1, 2... N,
 \end{equation}
where $d_n$, $S_n$, $Tr_n$ and $R_n$ means the data point, periodic component, trend component and residual component, respectively. \rv{Critical STL parameters for each historical data are (1) $n_p$, the number of observations per periodic cycle, which is \rv{$T_n$} for daily data and $24T_n$ for hourly data; (2) $n_i$, the number of passes of the inner loop, which is 1; (3) $n_o$, the number of passes of robustness iterations of the outer loop, which is 5; (4) $n_l$, the smoothing parameter of the low-pass filter, which is $[n_l]_{odd}$; (5) $n_s$, the smoothing parameter of the low-pass filter, which is set as 15; and (6) $n_t$, the trend smoothing parameter, which is $[1.67n_p]_{odd}$.} 

\subsection{Anomaly Scoring} After getting the estimated period value ($T_n$) for each historical data and the numeric value ($S_n$, $Tr_n$ and $R_n$) for the aforementioed three components, the respective anomaly scores (ASs) of three patterns are calculated in the anomaly scoring process based on their following detectors (as shown in Fig.~\ref{fig:algorithm}(2)), expressed as follows:
\begin{enumerate}[leftmargin=*,label={\textbf{({\arabic*})}}]
 \setlength{\itemsep}{2pt} 
 \setlength{\parskip}{2pt}
 \setlength{\parsep}{2pt}
 
 \item \textbf{Periodic:} 
  \vspace{-0.2cm}
 \begin{equation}
  \vspace{-0.05cm}
     AS_{periodic} = min\left(\frac{|T_n-T_{n-1}|}{T_{n-1}},1\right),
 \end{equation}
 where $T_n$ and $T_{n-1}$ denote the detected periods at the $n$th and $(n-1)$th timestamp at the periodicity detection step, respectively. As such, $AS_{periodic}$ describes the seasonal level shift. 
 
 \item \textbf{Trend:}
  \vspace{-0.05cm}
 \begin{equation}
  \vspace{-0.05cm}
     AS_{trend} = min\left(\left|\frac{K_n-K_{n-1}}{K_{n-1}}\right|,1\right),
 \end{equation}
 where $K_n$ is defined as the estimated slope of the trend component sequence $\textbf{Tr} = \{Tr_{n-L},Tr_{n-(L-1)},...,Tr_{n-1},Tr_n\}$ through linear regression, and $K_{n-1}$ represents the counterpart at the $(n-1)$th timestamp. Similarly, the trend level shift is indicated by $AS_{trend}$. 
 
 \item \textbf{Spike:}
  \vspace{-0.05cm}
 \begin{equation}
  \vspace{-0.05cm}
     AS_{spike} = min\left(\left|\frac{R_n-\mu_{n-1}-3\sigma_{n-1}}{3\sigma_{n-1}}\right|,1\right),
 \end{equation}
where $\mu_{n-1}$ and $\sigma_{n-1}$ denote the mean and variance with respect to the residue set $\textbf{R} = \{R_{n-L}, R_{n-(L-1)},...,R_{n-1}\}$. This metric is used for the evaluation of unexpected outlier.
\end{enumerate}
\vspace{-0.2cm}
\rv{We take the minimum differences as scores for equations (4--6) so to scale the anomaly score into [0,1], with 1 as the most anomalous situation. The final anomaly score at the $n$th timestamp can be calculated by
  \vspace{-0.05cm}
 \begin{equation}
  \vspace{-0.05cm}
     AS = f(AS_{periodic},AS_{trend},AS_{spike}),
 \end{equation}
where the aggregation function $f$ can be customized by various selectors, such as $min$, $max$, and different \textit{weighted average}, etc. Our algorithm is incremental; i.e., at every timestamp, it only utilizes the most recent data, rather than long-term data, to update results. Long-term data are not reasonable as the data distribution may be changed by various operations and human interventions. By using only the most recent data, our model can quickly accommodate itself to data distribution drifts. In addition, the AS indicates a node's anomaly degree at every timestamp so that engineers are able to rank the nodes efficiently and reduce their efforts in anomaly detection and diagnosis. In the future work, we need a more optimal method to aggregate the anomaly scores of three components. Moreover, the algorithm is only applied to one metric data now, so the correlations among different metrics of a node can be considered to incorporate in the process. }



\begin{figure*} [t]
 \centering 
 \includegraphics[width=\linewidth]{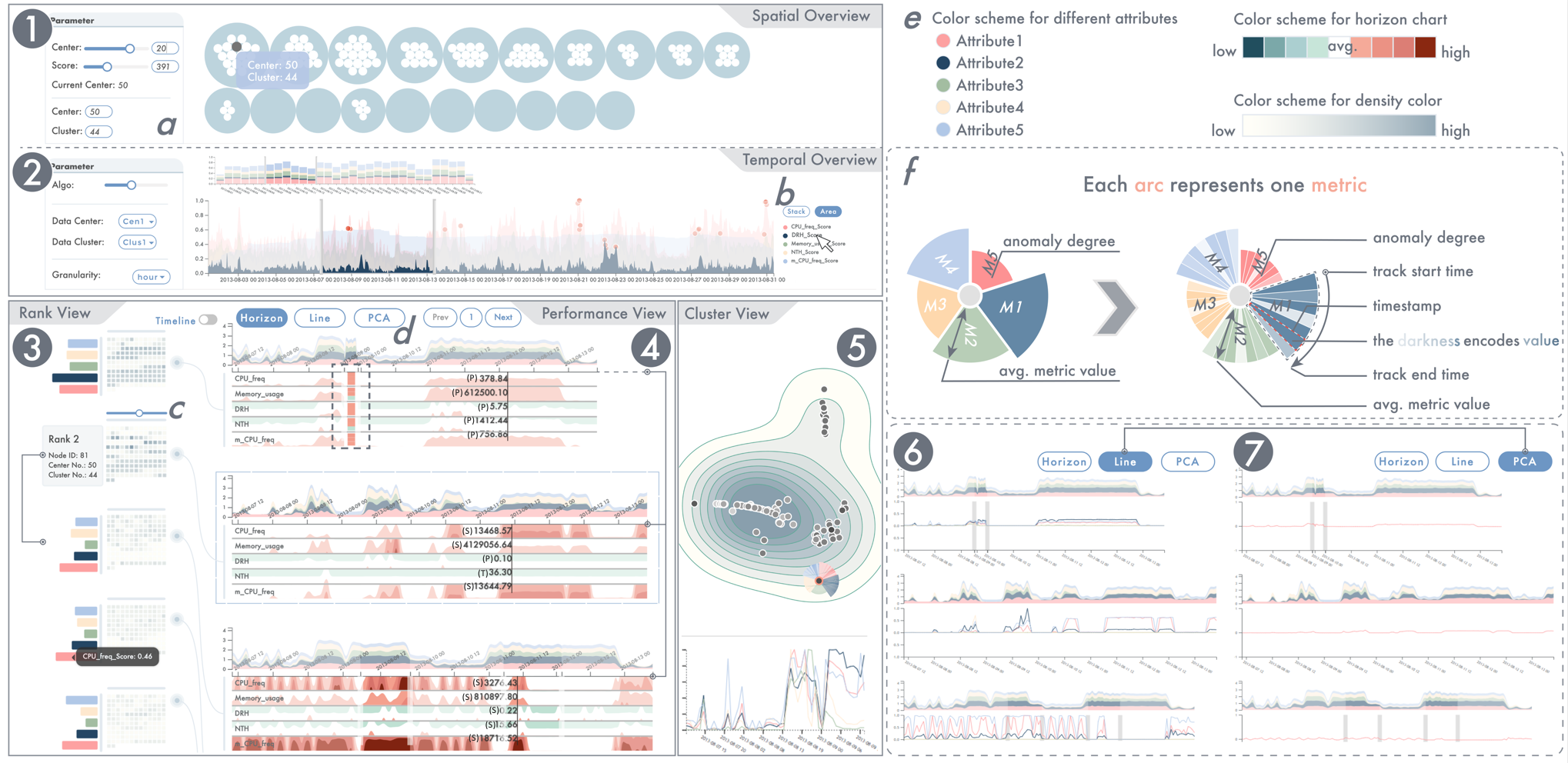}
 \vspace{-0.4cm}
 \caption{The \name system contains \rv{five interactive modules: (1) \textit{Spatial Overview}, (2) \textit{Temporal Overview}, (3) \textit{Rank View}, (4) \textit{Performance View} in \textit{the horizon chart} mode, and (5) \textit{Cluster View}. \textit{The performance view} contains two other views/modes: (6) \textit{the multi-line mode} and (7) \textit{the PCA mode}. Users can select and filter data in (a), switch to different visualization modes in different views by buttons (b) and (d), and change the layout by the slider bar in (c). (f) is the explanation of the glyph in (5). (e) shows the color schemes used in different views.}}
 \vspace{-0.4cm}
 \label{fig:interface}
\end{figure*}

%% file: sections/05-visual.tex
\section{Visualization}
\label{sec:visual}

This section presents the design considerations of visualization components derived from the discussions with our expert users \rv{in Section 5.1} and a brief summary of the interface \rv{in Section 5.2}, followed by the technical details of each component \rv{from Section 5.3 to 5.8}.

\subsection{Design Tasks}
A list of design tasks was formulated to guide the visualization designs based on the requirements outlined in \textbf{R1--R4}. We discussed with the \experts about the difficulties in determining the abnormal performance and the design ideas that are able to solve the problem. For example, the most commonly mentioned difficulty is to scale the large-size data and connect all the relevant information of the detected anomalies together to help \experts decide whether it is a true anomaly or not. The traditional methods usually generate too many false positive results. To solve these challenges, the experts came up with some initial ideas (e.g., the stacked line charts) for visualization. In general, they desired a tool that can facilitate the identification, interpretation, and reasoning of detected cloud computing anomalies, thereby improving the maintenance of cloud centers. Guided by these considerations, we decided on a list of visualization tasks for two main purposes: (1) facilitating the exploration of data and the identification of anomalies and (2) enhancing the rationalization of the detected anomalies. We summarize these design tasks as follows.

\begin{enumerate}[leftmargin=*,label={\textbf{T{\arabic*}}}]
 \setlength{\itemsep}{2pt}
 \setlength{\parskip}{2pt}
 \setlength{\parsep}{2pt}

\item \textbf{Show the overview of anomaly detection results for data query.} \rv{A large-scale cloud computing system usually contains multiple data centers, with each center containing hundreds of clusters that host tens of thousands of servers. Hence, it is critical to provide visualization techniques that can summarize the cloud computing performances and anomaly detection results to help experts narrow down the search space.}
\item \textbf{Rank the suspicious computing nodes dynamically.}
To reduce the search effort for suspicious nodes, visualization should be designed to aid the searching and filtering of anomalous performances by ranking the nodes whilst preserving the time context, e.g., displaying the periodic pattern of the operating activities.
\item \textbf{Browse the data flexibly in multiple ways.}
Despite the importance of the anomaly scores and ranks, \rv{the raw performance data that contain different metrics are of most concern for experts to identify an anomaly case.} Therefore, rich interaction techniques should be designed to enable an efficient mechanism to explore the performance data and extract their anomalous patterns.
\rv{\item \textbf{Facilitate anomaly detection and interpretation.} The visualization design of the system should consider the combination of the numeric anomaly detection results with the performance metric data. The visualization and interaction should enable users to make comparisons and display correlations over performance metrics at different levels from data summarization to focused context.}
\item \textbf{Display the similarity of different computing nodes.}
In addition to displaying the temporal patterns, another key to understanding the anomalous cloud computing performance is to differentiate anomalous compute nodes from normal ones. To this end, the system should show the clustering of the nodes based on their similarities, revealing some common features that led to the anomaly.

\end{enumerate}

\subsection{User Interface}
Guided by the above design tasks and the experts' feedback, we designed our visualization and interaction modules. The user interface (UI) of the \name system, as shown in Fig.~\ref{fig:interface}, consists of five modules: \textit{the spatial overview} (Fig.~\ref{fig:interface}(1)), indicating the general anomaly degree based on the cloud system hierarchy, from data centers to their sub-level data clusters; \textit{the temporal overview} (Fig.~\ref{fig:interface}(2)), displaying the anomaly distributions over time for data filtering (\textbf{T1}); \textit{the rank view} (Fig.~\ref{fig:interface}(3)), showing the highly suspicious computing nodes in the descending order of their anomaly scores (\textbf{T2}); \textit{the performance view} (Fig.~\ref{fig:interface}(4)), associating the anomaly detection results with the raw performance metrics data and facilitating the anomaly inspection and correlation analysis via three \rv{visualization modes (the other two modes are \textit{the multi-line mode} in Fig.~\ref{fig:interface}(6) and \textit{the PCA mode} in Fig.~\ref{fig:interface}(7))} (\textbf{T3--5}); \textit{the cluster view} (Fig.~\ref{fig:interface}(5)), summarizing the activities of all the computing nodes (both normal and abnormal) and clustering them based on their similarity terms of data features (\textbf{T6}). All the views are interactively connected, illustrating different contexts of a set of top suspicious computing nodes. Different color schemes are designed to illustrate the different information (Fig.~\ref{fig:interface}(e)). In particular, categorical colors are used to represent the information related to each of the performance metrics. A linear color scheme, ranging from white to grey, is used for indicating the anomaly score from low to high, while a dichotomous color scheme, ranging from dark blue, to white, to dark red, encodes the raw data values of the performance metrics, with red/blue encoding a value higher/lower than the average (white). More design details of each view are introduced in the following sections.

\begin{figure} [t]
 \centering 
 \includegraphics[width=\linewidth]{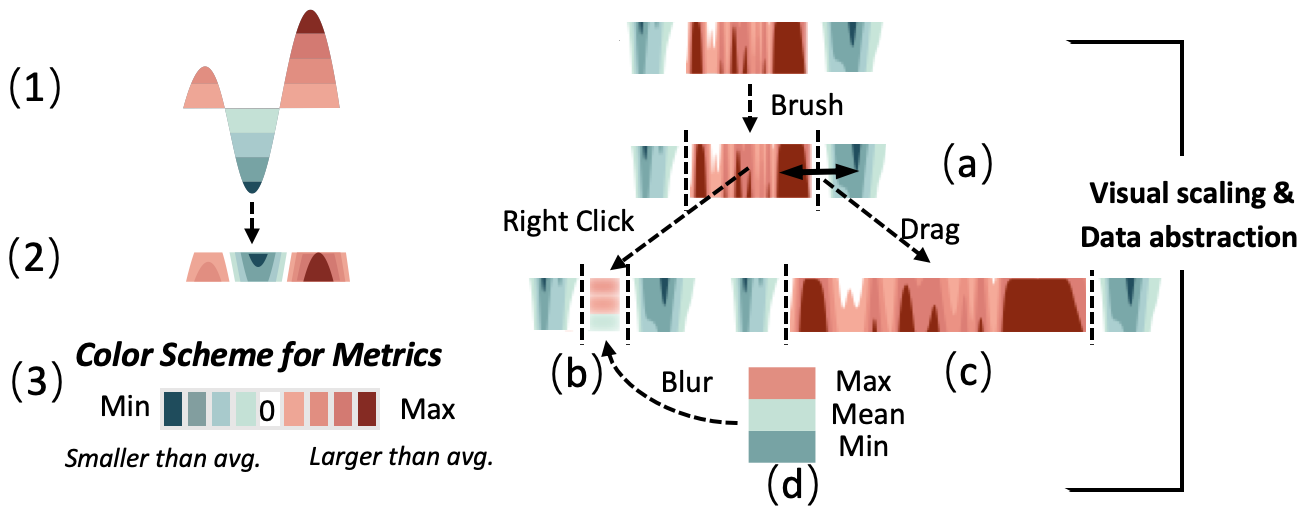}
 \vspace{-0.5cm}
 \caption{(1)--(3) show the standard horizon chart design \cite{heer2009sizing}. The horizon chart can be extended by: (a) selecting a specific period by brushing, and then (b) visually collapsing this period into a glyph for data abstraction and time scaling, or (c) stretching this period for detailed inspection.}
 \vspace{-0.4cm}
 \label{fig:view}
\end{figure}

\subsection{Initial Inspection and Data Selection}
The \name system contains two views for initial anomaly analysis and data query: (1) \textit{the spatial overview} displays the overall anomaly distribution based on the hierarchy of the cloud computing system; (2) \textit{the temporal overview} depicts the variance of the summarized anomaly score with time. Both of them allow users to select a more detailed data subset for in-depth anomaly inspection (\textbf{T1}).

\subsubsection{Spatial Overview}
\textit{The spatial overview} assists users to observe the general anomaly degree and query the data of a cloud computing system hierarchically (from data center to sub-level data cluster) through a bubble chart. Firstly, as shown in Fig ~\ref{fig:interface}(1), each blue outer bubble represents a data center and its interior white bubbles represent data clusters belonging to that center. According to the experts' feedback, abnormal instances are relatively rare and they hoped the system could directly show them the most likely anomalies. Thus, \textit{the spatial overview} arranges data centers according to their abnormal scores in descending order, from left to right and top to bottom. The bigger a bubble is, the larger its anomaly score is. The calculation of the anomaly score is to sum the scores of all nodes in that center. \rv{In addition, the inner bubbles will only appear when their represented data clusters' sum of anomaly scores is larger than the human-set threshold.} Both the number of data centers and the threshold value for data clusters can be set through two sliders (Fig.~\ref{fig:interface}(a)). Based on the displayed information, users can query the data from a specific data center and cluster in two ways: (1) clicking on a white bubble of interest (the tooltip will show the name of the corresponding data center and cluster) to select the tracking records from its represented data cluster; (2) inputting a specific data center and data cluster ID by using the input boxes.
\subsubsection{Temporal Overview}

\textit{The temporal overview} (Fig.~\ref{fig:interface}(2)) aims at revealing an anomaly overview of the variation of all the important tracking metrics (e.g., CPU frequency and memory usage) over time, which provides users with a general impression of the whole data set in a temporal context. We present two types of charts, namely, an area chart and a stacked bar chart, to show the \rv{sum of all nodes's anomaly scores in terms of different metrics at every timestamp}. The y-axis shows the anomaly score, and a categorical color scheme is used to represent different metrics. Specifically, the area chart overlaps the different metrics; thus the most anomalous metric at every time point can be highlighted, and the inter-pattern comparison and correlation discovery of multiple metrics can be fulfilled (\textbf{T5}). By contrast, the stacked bar chart emphasizes the total amount of anomaly scores for all the performance metrics at every time point. To assist users in catching the time period of interest more efficiently, we mark the top five points of each performance metric with red dots for reference. Different interaction designs are provided in this view: \rv{(1) brushing a time period in the overview for further analysis; (2) switching the view mode, as well as filtering out some metrics or highlighting others (tuning the corresponding color opaque) (Fig.~\ref{fig:interface}(b));} (3) choosing three types of time granularities, namely, minutes, hours and days, as an input parameter to show the performance history at different levels of detail (Fig.~\ref{fig:interface}(2)). 

\subsection{Anomaly Ranking}

\textit{The rank view} in Fig.~\ref{fig:interface}(3) shows a list of compute nodes with high anomaly scores within the user-specified time period in the temporal view, which reduces users’ efforts in searching for suspicious nodes. \rv{To rank the nodes, we sum the anomaly scores of different metrics (equation (7)) for each node at every timestamp, and the larger the sum, the more abnormal the node.} Each chart in this view represents one compute node and is placed in the increasing order of anomaly ranks, which consists of three components: an information card, an anomaly calendar view and a line connecting \textit{the rank view} with \textit{the performance view}. First, the information card employs a bar chart showing the average anomaly degree of different performance metrics. When clicking on this chart, the card will flip and provide textual information about the node, such as the node ID and the node rank. Second, the anomaly calendar view depicts the temporal pattern, especially the potential periodic patterns of a node. Each cell in this view represents one time unit according to the selected time granularity (minute, hour, day). The color, ranging from white to grey, encodes the sum of different metrics' anomaly score on one time unit, for example, it gets darker as the anomaly score increases. Various interaction techniques extend the functionality. Users can modify the arrangement of each calendar via tuning the slider bar for each node in Fig.~\ref{fig:interface}(c). Thus potential periodicity of different nodes may be observed. 
Finally, the lines between \textit{the rank view} and \textit{the performance view} establish a correspondence between the general information and the detailed performance metrics of each node. 
\rv{With the buttons on the right of Fig.~\ref{fig:interface}(d), the user can inspect different nodes according to their rankings (\textbf{T2}).}

\subsection{Performance View}

\textit{The performance view} (in Fig.~\ref{fig:interface}(4)) displays the detailed performance metrics data, associated with anomaly detection results, to facilitate the anomaly inspection and correlation analysis among different metrics in time scale. Multiple visualization modes and interaction designs are provided in this view to improve the analysis efficiency. The charts are placed in increasing order of anomaly ranking, with the gap between two charts meaning the difference of anomaly scores between two represented nodes. The visualization for each node is composed of two sub-views. On the top, a stacked area view shows different metrics' anomaly scores and their comparison in proportion over time. A consistent categorical color scheme is used in this part to denote different metrics. By observing the height of the graph, users can easily detect when an anomaly occurs and which metric is abnormal. Thus they can quickly transfer to the suspicious time periods and metrics in \textit{the performance view} for detailed analysis. On the lower half, there is a performance view that summarizes and displays the variation of different performance metrics data over time in three modes, allowing the understanding of correlations and temporal patterns with different semantics and scales. 

\textbf{Horizon Chart Mode.} The horizon chart uses a space-efficient method to clearly show the overall trend of each performance metric and improve the comparison among different metrics over time. Each layer represents one metric in this view. Fig ~\ref{fig:view}(1, 2) illustrates the construction of the standard horizon chart, which is extended from a normal line chart that displays the metric value changes over time. Specifically, we offset the negative values such that the zero point for the negative values is at the top of the chart.
The metric value is indicated by a dichotomous color scheme ranging from green to white to red (Fig.~\ref{fig:view}(3)), with green/red encoding a value higher/lower than the average (white). The average value for a given metric is calculated by considering all the nodes in the data cluster.
We can interpret the horizon chart with the following logic: (a) use the color to decide whether the data is positive or negative; (b) observe the darkest color to decide the general range of the data; (c) follow the height of the darkest part to understand the temporal trend of the metric value; \rv{and (d) read the textual information to learn the real value of the corresponding metric, and the dominant reason (three causes mentioned in Section 4.2) for the anomalies (\textbf{T3--4}).}

\textbf{Multi-line Mode.} \rv{A multi-line chart is provided as a more conventional and familiar visualization type for users, compared with the horizon chart, to facilitate their understanding of the data.} By clicking the ``line'' button shown in Fig.~\ref{fig:interface}(d), the mode of \textit{the performance view} will change to \textit{the multi-line mode} (Fig.~\ref{fig:interface}(6)), where each line presents one metric and the same categorical colors used in our system are applied. Also, each attribute's data is normalized and scaled to the range of [-1,1] \rv{because we need to make the units for different attributes consistent with the same y-axis} (\textbf{T4}).

\textbf{PCA Mode.} The PCA (principal component analysis) view depicts the major trends of a node's performance (Fig.~\ref{fig:interface}(7)).
\rv{The node's performance data, which is a multivariate time series with several metrics (e.g., CPU frequency and memory usage), are projected to a one-dimensional time series with PCA analysis. Thereby, we can reduce the number of variables and make important trends in the data directly accessible, providing qualitative high-level insights for cloud computing performance and anomaly inspection \cite{aigner2008visual} (\textbf{T4}).}

\textbf{Glyph Design.} Given the large scale of cloud computing data in time scale, a visual aggregation glyph, following the ``magnet'' metaphor (Fig.~\ref{fig:view}(b)), is designed to facilitate anomaly inspection via conducting data abstraction for irrelevant or uninteresting periods in \textit{the performance view}, as well as saving space in the time domain when users have to undertake analysis for a long period. After brushing a time period in \textit{the performance view} (Fig.~\ref{fig:case1}(a)), users can right click in the brushed area. Then this area will collapse into multiple glyphs (Fig.~\ref{fig:case1}(b5)), with each one representing one metric's data. Each glyph is trisected into three rectangles (Fig.~\ref{fig:view}(d)), and the color of the top, middle and bottom rectangles encodes the maximal, average and minimal values of the corresponding metric in the collapsed area. The color scheme is the same as that of the horizon chart (Fig.~\ref{fig:view}(3)). Moreover, the glyph filling color is blurred (Gaussian blur) to present uncertainty (std value) of the data abstraction for this area. In addition, users can also drag the border of the brushed area to narrow/broaden it, thus scaling \textit{the performance view} and focusing on the information of interest (Fig.~\ref{fig:view}(c)). Finally, there is a dynamic axis which can follow the move of the mouse to immediately provide the real metric value for information reference.

\begin{figure} [t]
 \centering 
 \includegraphics[width=\linewidth]{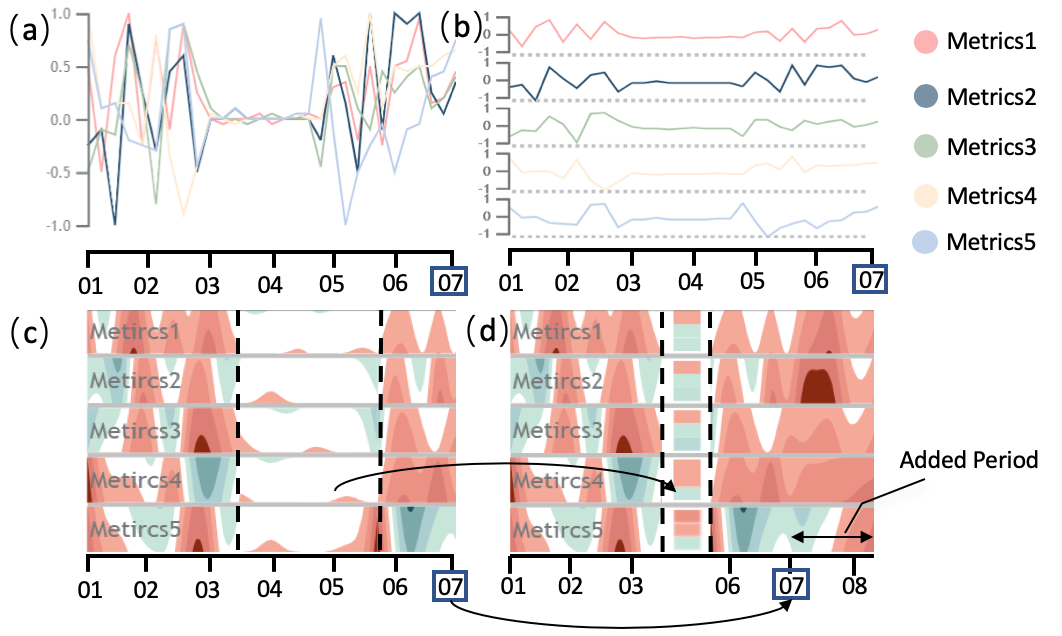}
 \vspace{-0.6cm}
 \caption{Alternative designs for performance metrics: (a) multi-line chart, (b) small multiples, (c) horizon chart. Our design (d) extends (c) with glyph and the nonlinear time scaling for analyzing more data.}
 \vspace{-0.5cm}
 \label{fig:alternative}
\end{figure}

\textbf{Alternative Design}
Several design alternatives for \textit{the performance view} were considered, as shown in Fig.~\ref{fig:alternative}. The first and most intuitive one was to use the multi-line chart containing all the metrics' performances in a single graph (Fig.~\ref{fig:alternative}(a)). However, this visualization has two problems: the severe visual clutter when data is large, and the inconsistency in measurement units for the different metrics. Another alternative was to draw line charts separately (Fig.~\ref{fig:alternative}(b)) for each metric and to align them horizontally. But this method consumes much more space for a single line chart if we want it to express the information as clearly as the horizon chart (Fig.~\ref{fig:alternative}(c)). The two methods both lack the ability to show the trends of multiple metrics clearly as horizon chart. However, a direct application of horizon chart does not support the scaling in time domain. Users would have difficulty in analyzing the overall changes or focusing on a detailed time period when the length of time series is very long. In this sense, the visual aggregation (glyph) and data abstraction techniques applied in the horizon chart enhance it by a nonlinear time scaling (Fig.~\ref{fig:alternative}(d)).

\subsection{Cluster View}
\rv{\textit{The cluster view} (Fig.~\ref{fig:interface}(5)) shows the spatial distribution of all the selected compute nodes using t-Distributed Stochastic Neighbor Embedding (t-SNE) based on their performance data, which are a multivariate time series. For each node, we construct a vector:
\vspace{-0.2cm}
\begin{displaymath}
\vspace{-0.2cm}
F=[m_{1,1},m_{2,1},...,m_{r,1},m_{1,2},m_{2,2},...,m_{r,2},..., m_{1,n},m_{2,n},...,m_{r,n}],
\end{displaymath}
where $m_{r,n}$ means the $r$th metric value at the $n$th timestamp. Therefore, we conduct the t-SNE dimensionality reduction based on these feature vectors. The clustering analysis can aggregate the multivariate temporal data into subsets exhibiting a certain similarity.} Each circular glyph in this view represents one compute node whose anomaly degree is encoded by the filling color of its inner dot, ranging from white to grey. The darker the color, the more abnormal it is. The anomaly degree is calculated by the Local Outlier Factor~(LOF) and normalized to -1 (normal) and 1 (abnormal). To enhance the anomaly and clustering analysis, we render a contour map to reveal areas with different densities, which are indicated by the filling color from white (low density) to grey (high density). Intuitively, a normal node tends to lie in high-density areas where many others have similar behaviors.
We use the kernel density estimation (KDE) to define the density at the user $u$'s position ($x_u$), which is formulated as
\vspace{-0.2cm}
\begin{displaymath}
\vspace{-0.2cm}
f(x_u)=\frac{1}{nh}\sum_{1}^{n}K(\frac{x_u - x_i}{h}),
\end{displaymath}
where $K$ is the kernel function, $x_i (i \neq u)$ indicates the positions of the nodes and $h$ is the kernel's bandwidth. 
From this view, we can also provide another perspective for anomaly diagnosis according to the inconsistent measurements of similar nodes. For example, some anomalous nodes (high LOF score) may be grouped in the white contour, which could represent a rare category that encounters the same problem (like network attacks) in cloud data centers. 
Furthermore, we design a glyph (Fig.~\ref{fig:interface}(f)), following the ``fan'' metaphor, to facilitate comparison among different nodes by summarizing their general usage metrics (\textbf{T6}).

\textbf{Glyph Design.} This glyph contains several arcs which are equally segmented around the inner circle. The left one in Fig.~\ref{fig:interface}(f) is our original design. Each arc illustrates one performance metric that can be denoted by the categorical colors that are consistently used in our system. The radius of the arc encodes the average value of its represented metrics \rv{across the selected time period}. However, it loses the \rv{temporal information} for each metric, so we redesigned the visualization \rv{as shown on the right in the figure}. We enhanced each arc with a circular timeline in clockwise order. For example, if a node contains eight records for CPU frequency in the selected time period, \rv{then the corresponding arc is divided vertically into eight equal segments, with each segment representing one record.} The lighter (closer to white) the segment is, the lower the record value is. Furthermore, users can click on a specific glyph to get the detailed raw metric data of the node at the bottom of this view, which is the same the multi-line chart employed in \textit{the performance view}. In this way, \rv{we can help users immediately get the low-level but detailed performance data of a node of interest, as well as connect the visualization results of the horizon chart.}

\subsection{Interactions}
The following interactions are designed to help with the data exploration and anomaly inspection.
\textbf{Anomaly Inspection.} Users can compare different levels of cloud computing data with different scales in the spatial and temporal overview. They can right/double click and zoom the visualizations in \textit{the performance view} to make a detailed inspection. A novel glyph summarizing a node's performances can be display or not in \textit{the cluster view}. \textbf{Query and Filtering.}
Users can load different subsets of data by using the query module or direct interaction on the views in Fig.~\ref{fig:interface}(1, 2), as well as focus on a specific metric in \textit{the temporal overview} by clicking the legend on Fig.~\ref{fig:interface}(b) (\textbf{T1}).
\textbf{Switching Context.} Users can switch between different visualization modes in \textit{the performance view}. In \textit{the temporal overview}, they can choose the area chart or stacked bar chart, and change the time granularity for all the corresponding views based on the analytic requirements.
\textbf{Zooming and Scaling.}
Four views support zooming for exploring a large set of data items, namely, \textit{the spatial overview}, the calendar graph in \textit{the rank view}. Particularly, users can brush a period of time for the node in \textit{the performance view} and choose to expand it for more details or narrow it for information reduction.
\textbf{Data Abstraction.}
We design two novel glyphs to generalize different data information for data exploration and anomaly inspection, namely, the ``magnet'' in \textit{the performance view} and the ``fan'' glyph  in \textit{the cluster view}.
\textbf{Tooltips and Highlighting.}
Tooltips are provided in \textit{the spatial overview}, \textit{the rank view} and \textit{the cluster view} for more reference information. In addition, there is a dynamic axis in \textit{the performance view}, following the moving of the mouse, to provide the raw value and dominant anomaly pattern of the performance data in real time. All the views are linked together. 

%% file: sections/06-evaluation.tex
\section{Evaluation}
\label{sec:eval}
We evaluated the effectiveness of \name via a quantitative performance evaluation of the proposed algorithm \rv{in Section 6.1}, two case studies with real-world data in \rv{Section 6.2}, and follow-up interviews with domain experts \rv{in Section 6.3}.

\begin{figure} [t]
 \centering 
 \includegraphics[width=\linewidth]{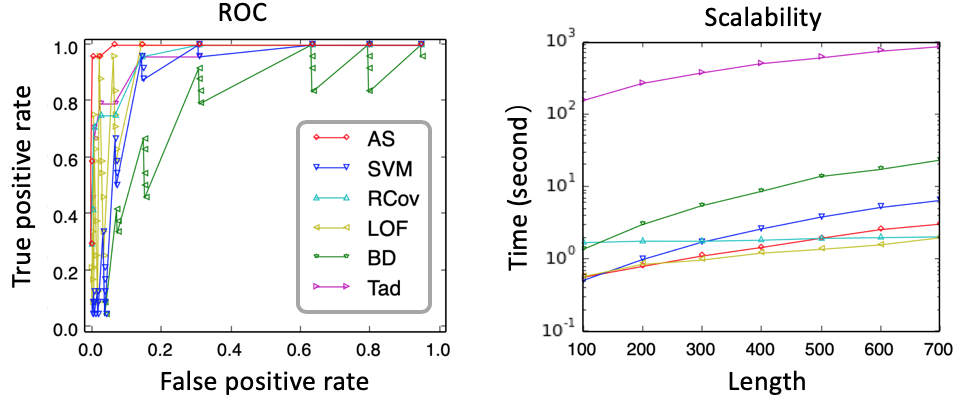}
 \vspace{-0.5cm}
 \caption{Algorithm performance evaluation. Left: the result indicates that our algorithm (AS) outperforms others in accuracy. Right: our algorithm can scale and exhibit linearly with the varying length of time series. }
 \vspace{-0.5cm}
 \label{fig:algoper}
\end{figure}

\subsection{Quantitative Evaluation}
We evaluated the effectiveness and scalability of the proposed algorithm through a quantitative comparison with five baseline methods: \rv{One-Class SVM \cite{chang2011libsvm}, Robust Covariance Estimation (RCov), LOF \cite{breunig2000lof}, Bitmap Detector (BD) \cite{wei2005assumption} and Twitter anomaly detection (Tad). They are selected as the most typical algorithms from five major categories of anomaly detection techniques, i.e., classification-based, neighbor-based, statistic-based, SAX (symbolic aggregate approximation)-based and S-H-ESD (Seasonal Hybrid ESD)-based methods, respectively. Other anomaly detection algorithms, like iForest \cite{liu2008isolation}, were discarded as their execution times are much longer than that of our proposed algorithm. We used the implementation of these models in the scikit-learn \cite{pedregosa2011scikit}, luminal and Twitter anomaly detection package in Python.} 

\textbf{Accuracy.} \rv{The standard information retrieval metric ROC was used here for accuracy evaluation due to the extremely imbalanced ratio between positive and negative instances. The accuracy of the algorithm was verified using real\_31.csv from the labeled Yahoo! S5 real time-series dataset \cite{yahoos5}. This dataset contains mixed change patterns (both long-term and short-spike anomalies) with a 1.68\% outlier percentage (24/1427), which can show the superiority of our algorithm better than using the dataset with a single anomaly pattern. We tested the accuracy of each algorithm with ten different values of its chosen parameter. In particular, our proposed algorithm took the number of recent "historical data" $L$ as the adjustable parameter (see Section 4.1), which grew from 5 to 50, with 5 as the step length. Then we ran each parameter setting of the algorithm ten times based on different discrimination thresholds (proportion of anomalies) [0.005, 0.01, 0.02, 0.04, 0.08, 0.16, 0.32, 0.64, 0.8, 0.95]. The thresholds grew exponentially, except for the last two. Therefore, we obtained 100 runs (10 parameter values $\times$ 10 discrimination thresholds) for each algorithm in the accuracy evaluation, which reduced the bias of detection results to different datasets. (Please refer to the supplementary material for more algorithm setting information and evaluation results. \footnote{\protect\label{supp}\url{https://lukeluker.github.io/supp/vast19\_cloudDet.pdf}})}

As shown in Fig.~\ref{fig:algoper}. Overall, our algorithm (denoted as ``AS'') outperformed the five baseline methods when there are different anomaly patterns in data. In particular, our algorithm had a higher true positive rate when the false positive rate was low (below 0.2), which means that our algorithm can reduce false positive anomalies when detecting the same number of anomalies as the baseline algorithms. This is important for cloud computing anomaly detection because the experts can save effort in anomaly diagnosis when the data scale is very large.

\textbf{Scalability.} \rv{The proposed algorithm must be scale-out and efficient to cope with large-scale data generated by cloud computing systems. To perform the evaluation, we tested the execution time of each algorithm by varying the length of the input time-series data. Specifically, in the experiments, each algorithm was executed on 50 time-series datasets (real\_1 to real\_50), with each dataset running ten times due to ten values for its adjustable parameters. We only varied the length of the input time series in different experiments, growing from 100 to 700 points with 100 as step length, to determine the execution time of the algorithm. The length was changed by selecting the first 100--700 data points of a dataset.} The experiments aere conducted in an eight-core (Intel Core i7-4790  CPU@3.6GHz) Window 10 computer with 16 GB memory. The results are summarized in Fig.~\ref{fig:algoper}. The figure suggests that the execution time of our algorithm can scale and exhibits linearly with the length of the time series. ``AS'' is less scalable than RCov and LOF, probably due to the time spent for pattern matching before anomaly detection, but the difference is acceptable considering the higher accuracy in detection results compared with other algorithms. \rv{Although Tad performs well in accuracy test in general, it has a worst computing speed compared with others.}

\rv{\textbf{Limitations.} Although the results show that our proposed algorithm has sound performance when considering the speed and accuracy together, the validity of the results needs further evaluation. There exist some factors that may affect the validity, such as the selection of parameters and their settings, the bias in the tested datasets and the insufficiency of the evaluation metrics. In particular, our algorithm may perform worse than the baselines when the anomalies merely have specific patterns like short-term spikes. Therefore, the results of the proposed algorithm could be affected by the aggregation strategy of the anomaly scores from the three components. A more optimal or automated aggregation method should be developed future work.\textsuperscript{\ref{supp}}}

\subsection{Case Study \& Expert Interview}
We demonstrate the usefulness and usability of \name by analyzing two real-world data sets. The study was conducted in a semi-structured interview with two domain experts for two purposes: to provide real abnormal cases to showcase the effectiveness of our system, and to give user feedback on the system design.

\textbf{Study Set-up and Interview Process.} We invited two experts who have a high level of experience in cloud computing related domains to be interviewed in the form of case study. The one for case study 1 is a data scientist from a company's cloud computing service team, who is responsible for the cloud system analysis and maintenance (\ea). The expert for case study 2 is a researcher from the same company's data analytics group, and his main focus recently has been anomaly detection with cloud computing data (\eb). Both of them had no prior knowledge about the data used in the case studies and \rv{were not collaborators on our project}. Each interview lasted approximately 1.5 hours, with 30 minutes for the introduction of the system and 60 minutes for data exploration and anomaly detection by the experts themselves using our system. Notes and expert feedback were recorded in the process.

\textbf{Case Study 1: Bitbrains Datacenter Traces.} In the first case study, \rv{we used a one-month (2013/08/01 -- 2013/08/31) dataset from Bitbrains \cite{Bitbrains}, which contains the performance metrics of 500 VMs from a distributed data center.} We resampled the data into hourly granularity and selected five typical metrics for the study, namely, average CPU rate (MHz), maximum CPU rate (MHz), memory usage (KB), disk read throughput (KB/s) and network transmitted throughput (KB/s). After loading the data into the system, the expert, \ea, started by choosing the hourly granularity and observing \textit{the temporal overview} in the area chart mode (Fig.~\ref{fig:interface}(2)). Then he noticed the first time period in which many anomalous points were marked (\textbf{T1}), so he brushed this period (from Aug.~7, 12 am to Aug.~15, 12 am) to make a further inspection. 
The most abnormal node, compared with others via the horizon chart (Fig.~\ref{fig:case1}(b1)), was not only abnormal in the drastic spikes (both short- and long-term) in the memory and CPU usage, but also quite sporadic and irregular in the spike occurrence time. Moreover, there existed inconsistency between the network throughput and other metrics (\textbf{T5}). After knowing that the data were collected from cloud service for business computation for enterprises, he believed that this node was very likely to be abnormal due to the unusual operations caused by users. 

Another type of anomalous periodic performance was also noticed by \ea. Different from the above case, this node, as shown in Fig.~\ref{fig:case1}(b2), had a periodic pattern in CPU usage while all the other metrics were stable. Then the expert tuned the slider bar at \textit{the rank view} and found the cycle was about 12 hours (Fig.~\ref{fig:case1}(a2)), which could be caused by regular analytic tasks run by the system customers
However, there was a sudden increase in all metrics from Aug.~11 to Aug.~13 (Fig.~\ref{fig:case1}(b2)), then felling to the same periodic performance as before. Similarly, when the expert switched to the next five nodes, he found an instance that had many near-periodic and short-term spikes at the beginning but an abnormal long-term increase near the end (Fig.~\ref{fig:case1}(b3)). To check whether this node returned to its previous behaviors after the anomalous periods, he brushed this period and right clicked it to transfer it to the ``magnet'' glyph for space scaling. Then the horizon chart was updated and the data after this period were shown (Fig.~\ref{fig:case1}(b5)). As expected, the node returned to similar periodic behavior as before. The expert also inspected the glyph and found that the part representing CPU frequency was blurred heavily, which means that this abnormal period was mainly unstable in CPU usage. Finally, when navigating in \textit{the performance view}, the expert found a node which seemed quite stable and normal with no spikes in the horizon chart (Fig.~\ref{fig:case1}(b4)), while its anomaly ranking was high (10th). Then he chose \textit{the PCA mode} to see whether there was any general trend in the data (Fig.~\ref{fig:case1}(b6)). \textit{The PCA mode}, however, showed a clear periodic pattern in the performance.

Finally, the expert selected other time periods via \textit{the temporal overview} and found some interesting patterns in \textit{the cluster view} (Fig.~\ref{fig:case1}(c)). For example, the distribution of nodes (Aug.6--17) showed three different clusters of nodes based on their behaviors. 
To compare the three clusters, the expert clicked some represented nodes in each cluster and obvious differences were observed. Compared with the normal nodes, one abnormal cluster (top-right) had a large red arc whose color was not distributed consistently, which means the highly changed CPU metric was the dominant reason for anomaly. By contrast, the other rare category had both large red and blue arc, which means they were abnormal due to their high value of several metrics. The expert thought that the nodes in the same cluster may perform similar tasks or belong to the same servers in the cloud computing systems (\textbf{T6}).

\begin{figure} [t]
 \centering 
 \includegraphics[width=\linewidth]{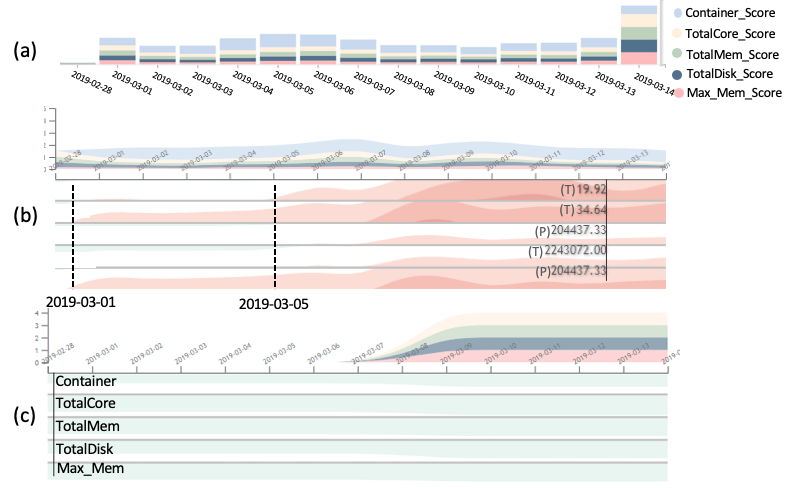}
 \vspace{-0.7cm}
 \caption{Identified anomalous patterns with live cloud system data.}
 \vspace{-0.4cm}
 \label{fig:case2}
\end{figure}

\textbf{Case Study 2: Live Cloud System Data.} \rv{We conducted the second case study with live cloud computing data (provided by the collaborator on the project) collected from about 1,000,000 nodes in the previous two weeks (2019/03/01 -- 2019/03/14), and we processed these time series into hourly resolution. In particular, the data were collected from more than 100 data centers, with each center containing about 20 data clusters, and each data cluster containing about 500 compute nodes. The performance metrics of the dataset were all about general resource allocations of a compute node, including container (VM) count, total core (CPU), total disk, average memory and maximum memory.} 
\eb, started by using \textit{the spatial overview} to select a data cluster for detailed inspection (\textbf{T1}). In this view (Fig.~\ref{fig:interface})(1)), he visualized the top 20 anomalous data centers and chose one anomalous data cluster from the first data center (\textbf{T2}). Then the visualization switched to \textit{the temporal overview} (Fig.~\ref{fig:case2})(a)). Considering that the resource allocation usually does not change frequently, he decided to conduct anomaly analysis with day granularity. The expert noticed a sudden increase in anomaly scores at the last day by observing \textit{the temporal overview} (Fig.~\ref{fig:case2})(a)), which means that special operations might have been undertaken on that day. Then he switched to \textit{the rank view} and \textit{the performance view} (\textbf{T2, 4}). After a brief inspection of the top-ranked notes, he quickly noted some nodes, like Fig.~\ref{fig:case2}(c), that have few changes in all resource allocations during the previous two weeks. He thought this was abnormal for a live system, and conjectured a physical maintenance or system breakdown for these nodes. In addition, when he compared some top-ranked nodes, he found another common issue that some nodes' metrics did not have synchronized behaviors like normal nodes. For example, starting from 03/01 in Fig.\ref{fig:case2}(b), the core and memory increase whilst the container number remains unchanged, which indicates that the resources allocated for each container increased. Moreover, around 03/05, there was a large increase in container number but a small increase in other metrics, which could be caused by adding some small containers into the node (\textbf{T5}). Such evolution in a compute node was abnormal due to the frequent inconsistent adjustment in the resource allocation compared with others in the same data cluster.

\subsection{User Feedback}
The experts from the case studies (\ea, \eb) provided a wealth of insightful feedback, which are summarized into three categories.

\textbf{Automated Anomaly Detection.} 
The experts showed a transition in attitude towards the automated anomaly detection algorithm. Before the case studies, neither of them trusted the algorithm because the traditional methods generated a large number of false positives in their past experiences. \ea stated, ``We don't have time to check the results one by one... even a few percentages of wrong [false positive] results for cloud computing is large in number.'' However, after inspecting the top-ranked results in \textit{the rank view} and \textit{the performance view}, both experts placed increased trust in the algorithm. For example, they commented that ``the algorithm can detect anomalies showing different anomalous patterns.''
Inspired by the PCA analysis for different metrics in our system, \eb further suggested that ``the algorithm could be more powerful if it can make a comprehensive anomaly detection by considering different metrics with different time granularity.'' 

\textbf{System.}
Both experts regarded the system as useful and user-friendly for analyzing cloud computing anomalies. ``It's useful!'' was the most common refrain from \ea, and he felt the findings were exactly the sort of information that cloud computing systems would use in anomaly analysis. ``It would be useful for maintenance and resource allocation of cloud computing systems,'' he said, ``We usually implement some resource allocation algorithms to the cloud system, but have difficulty in evaluating them... This tool can monitor our system and identify warnings [anomalies] from performance data.'' Moreover, \eb commended the workflow of the system as clear and logical because the system is consistent with the conventional way to analyze cloud system data.
However, both experts felt the system ``too comprehensive'' and ``a little overwhelming'' at the beginning of use. Although a short introduction was sufficient to bootstrap their analyses, \eb suggested that a tacit tutorial be included to guide independent users, and that the system could also display the information based on an exploration-based approach.

\textbf{Visualization and Interaction.}
Most visualization and interaction designs were thought to meet the design tasks. Both experts stated that ``the performance view is the essential part'' because it displayed ``the overall trend'' of tracking metrics and also supported ``a clear comparison'' among different metrics and nodes. \eb stated, ``I used to take the line chart to analyze the results... It's quite chaotic when there are too many metrics or long time periods.'' \ea mentioned the novelty of showing the similarity among compute nodes in \textit{the cluster view}. He said, ``I never thought of this,'' and ``It provides a new perspective to understand cloud computing anomalies.'' \eb thought the glyph in the horizon chart as ``interesting and useful'' because it can ``put aside uninterested information and save space for more data.'' Moreover, they regarded the interactions in our system as helpful, which allowed a quick way to navigate and scale the visualizations when needed.

%% file: sections/07-conclusion.tex
\section{Conclusion and Future Work}
\label{sec:conclusion}

We have presented \name, an interactive visualization system that enables expert users to identify and analyze anomalous performances in cloud computing systems. The system supports the exploration process in both spatial and temporal context by incorporating a novel anomaly detection algorithm, multiple coordinated contextual views and rich interaction designs. We demonstrated the power of \name through a quantitative evaluation for algorithm performance, two case studies using real-world data and follow-up interviews with domain experts. While these results are promising, there are several new directions for future work. \rv{First, the algorithm can still be optimized in many aspects like the aggregation strategy for different components' anomaly scores.} Second, we plan to deploy our system on cloud computing platforms so as to discover ways to improve it through more user feedback. \rv{To improve the qualitative evaluation, we intend to compare \name with established application monitoring (APM) tools used by cloud operators and implement it with more datasets. To turn the approach into a supervised one,} we can tune the anomaly detection model responsively by feeding user corrections (e.g., re-rankings and anomaly labeling) back to an active learning process in real-time. Finally, we will evaluate the usability of our system and glyph designs via a formal user study \rv{such that simplifying our system by removing the uncritical parts}.


\balance

%% file: sections/09-acknowledgement.tex
\section{Acknowledgements}
We would like to thank all the reviewers and domain experts for their comments. This work is partially supported by a grant from MSRA (code: MRA19EG02).